\documentclass[10pt,aps,onecolumn, preprint,superscriptaddress]{revtex4-1}
\usepackage[utf8]{inputenc}
\usepackage[pdftex]{graphicx}
\usepackage{hyperref}
\hypersetup{
    colorlinks=true,       % false: boxed links; true: colored links
    linkcolor=red,          % color of internal links (change box color with linkbordercolor)
    citecolor=blue,        % color of links to bibliography
    filecolor=magenta,      % color of file links
    urlcolor=cyan           % color of external links
}

\usepackage{url}
\usepackage{mathtools}
\DeclarePairedDelimiter{\evdel}{\langle}{\rangle}

\begin{document}
\title{Some minor insights into on-demand ride-sharing}
\title{Network topology and on-demand ride-sharing}
\title{Topology dependence of on-demand ride-sharing}

\author{Debsankha Manik}
 %\email[Electronic Address: ]{molkenthin@nld.ds.mpg.de}
 \affiliation{Max Planck Institute for Dynamics and Self-Organization (MPIDS), 37077 Göttingen, Germany}
 
\author{Nora Molkenthin}
 %\email[Electronic Address: ]{molkenthin@nld.ds.mpg.de}
 \affiliation{Max Planck Institute for Dynamics and Self-Organization (MPIDS), 37077 Göttingen, Germany}

\affiliation{Chair for Network Dynamics, Institute for Theoretical Physics and Center for Advancing Electronics Dresden (cfaed), Technical University of Dresden, 01069 Dresden}

\begin{abstract}
\textbf{Abstract:} Traffic is a challenge in rural and urban areas alike with negative effects ranging from congestion to air pollution. Ride-sharing poses an appealing alternative to personal cars, combining the traffic-reducing ride bundling of public transport with much of the flexibility and comfort of personal cars. Here we study the effects of the underlying street network topology on the viability of ride bundling analytically and in simulations. 
Using numerical and analytical approaches we find that system performance can be measured in the number of scheduled stops per vehicle. Its scaling with the request rate is approximately linear and the slope, that depends on the network topology, is a measure of the ease of ridesharing in that topology.  This dependence is caused by the different growth of the route volume, which we compute analytically for the simplest networks served by a single vehicle.
\end{abstract}

\maketitle

\section{Introduction}
The increasing demand for mobility in modern urban, suburban and rural areas presents a wide range of ecological and logistic challenges. While urban areas struggle with traffic jams, air pollution and parking space shortages \cite{eu2018_airQualityReport,nyc2018_mobilityReport}, rural areas are often unable to provide accessible and frequent public transport. 
The recent rise of the sharing economy \cite{belk2014you, cohen2014ride, kamargianni2016critical,greenblatt2015automated} has brought up ride-sharing as a possible answer to all of these problems.
Ride-sharing poses an appealing alternative to personal cars, combining the traffic-reducing ride bundling of public transport with much of the flexibility and comfort of personal cars \cite{spieser2014toward,zhang2016control,barbosa2018_mobility_models,humancities,vazifeh2018_minimumFleetProblem}. 
Intelligent on-demand ride-sharing services are hoped to reduce the ecological footprint associated with individual mobility by dynamically bundling rides together, reducing the amount of vehicles necessary for the same number of rides \cite{tachet2017scaling, Santi2017,sorge2015towards,sorge2017towards}.

However, the complex behaviour of such dynamic dial-a-ride problems (DARP) \cite{berbeglia2010dynamic} is not yet fully understood. Recent studies have examined the dynamical behaviour of specific ride-sharing strategies analytically \cite{herminghaus2019mean} or in simulations \cite{alonso2017demand,ma2013t,agatz2011dynamic,horn2002fleet}. However, the general scaling behaviour, or dependence on street network topology and request patterns are not currently understood. Such an understanding would be necessary to compare different dispatching strategies and network settings and make informed decisions about which dispatching strategy works best for a particular network.

Here we study the effects of the underlying street network topology on the viability of ride bundling analytically and in simulations in the low-density limit by studying the performance of a single vehicle. We find that for finite request rates and vehicles not restricted by capacity, there is always a quasi-stationary regime of operation, varying in waiting time and typical vehicle occupancy. We develop a probabilistic description of the steady-state route length, relying on \emph{route-volume}, a topological characteristic of a network that we define, and use this to derive the equilibrium stop-list length. Based on this we show the scaling of the steady-state stop-list length with the dimensionless request density to be linear with a slope depending on the topology. The dependence of the route-volume on the stop-list length can be approximated explicitly for some simple topologies (ring and star) and numerically otherwise.

We apply this analysis to unweighted real-world street networks. The general layout of urban centers is predominantly grid-like in structure, whereas rural areas appear to be best described as interconnected rings with long stretches of unbranching streets. This leads to the surprising effect that, while the request density of cities tends to be better suited for ride-sharing, the topologies show the opposite trend with rural areas allowing easier bundling. This is particularly important as cities already have well functioning public transport options, which have proven to be impractical in less densely populated areas.

\section{Model}
\label{sec:model}
In order to eliminate non-topological influences on ride-sharing, we reduce the system to a simple graph with $N$ nodes and a request pattern $P_{i,j}$, serviced by a \emph{single} vehicle. A request is an ordered pair of a pick-up node $i$ and a drop-off node $j$ drawn from the request pattern. New requests arrive according to a Poisson process with an average time $\Delta t$ between requests to be included in the route according to a \emph{dispatcher algorithm}. 
The dispatcher algorithm checks if the request's pick-up node can be inserted to the existing stop list without incurring any detour. If this is the case it checks if the drop-off node can also be inserted to the existing stop list without any detour, otherwise it is appended at the end of the stop list. If the pickup cannot be inserted with zero detour, then both the pick-up and the drop-off are appended right after each other at the end of the stop list.

We introduce the dimensionless request rate x in order to compare system properties across network topologies
\begin{equation}
 x = \frac{2\evdel{l}}{\nu \Delta t}, 
 \label{eq.x}
\end{equation}
where $\evdel{l}$ is the average length of the requested ride and $\nu$ is the bus speed.

A request rate of $x=1$ means that the vehicle covers a distance of $2\evdel{l}$ between two requests, where $2 \evdel{l}$ is the expected distance from the endpoint of the route to the new pick-up and from there to the new drop-off. Therefore even without ride-sharing, rides are on average completed within one $\Delta t$ and even a taxi system would be able to serve them all, operating at maximum capacity. For $x > 1$ the taxi can no longer serve the system and waiting times diverge, in this case the ride-sharing system transports $x$ times as many passengers as the taxi could. 

To quantify passenger satisfaction in a ride-sharing system, we investigate the service time $t_s$, i.e. the time it takes from placing the request until being delivered at the requested drop-off location. The number of planned stops in the system, the stop list length $n$, on the other hand serves as a measure of performance from the perspective of the system as a whole.

\begin{figure}
    \centering
    \includegraphics[width=0.3\columnwidth]{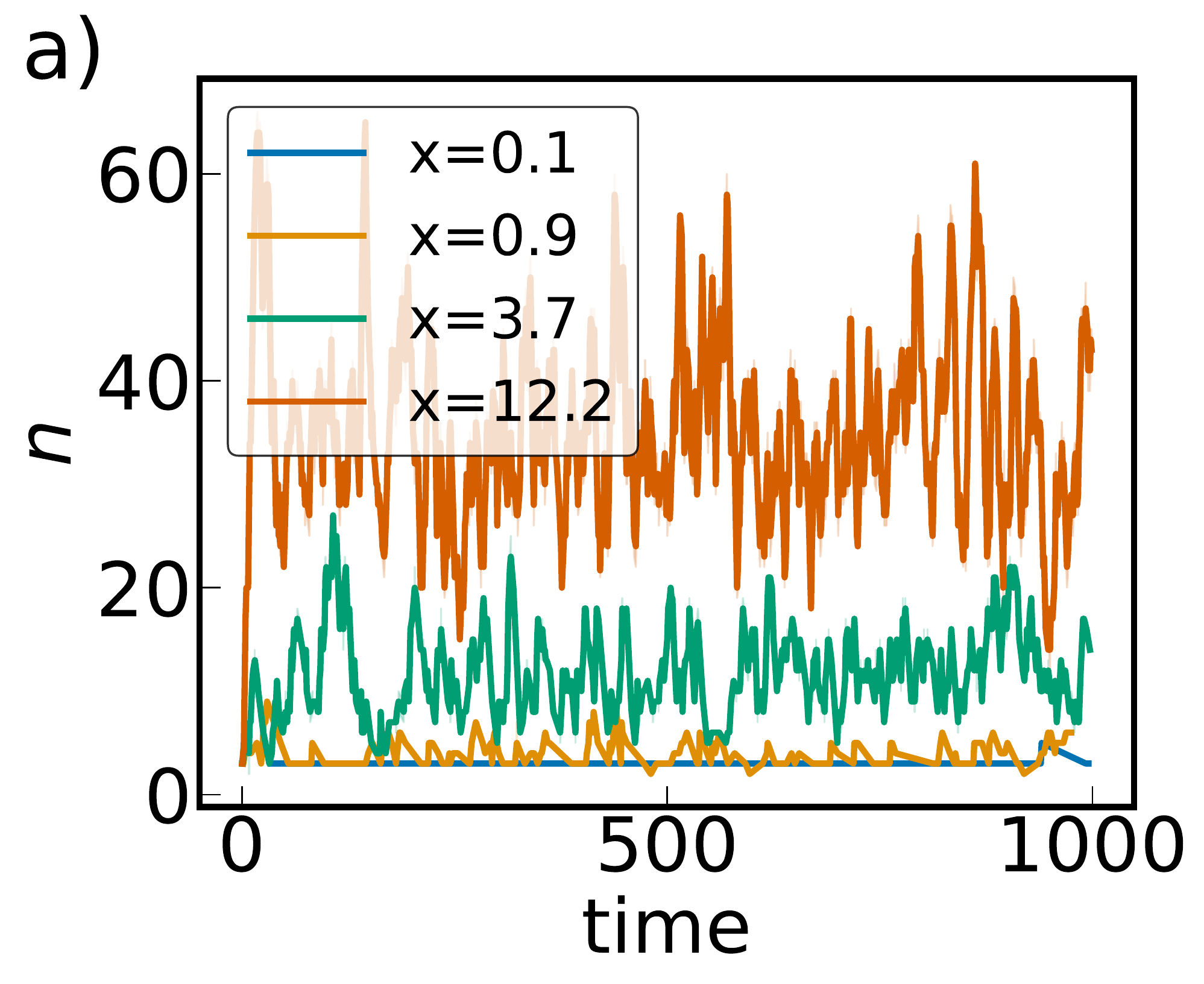}
    \includegraphics[width=0.3\columnwidth]{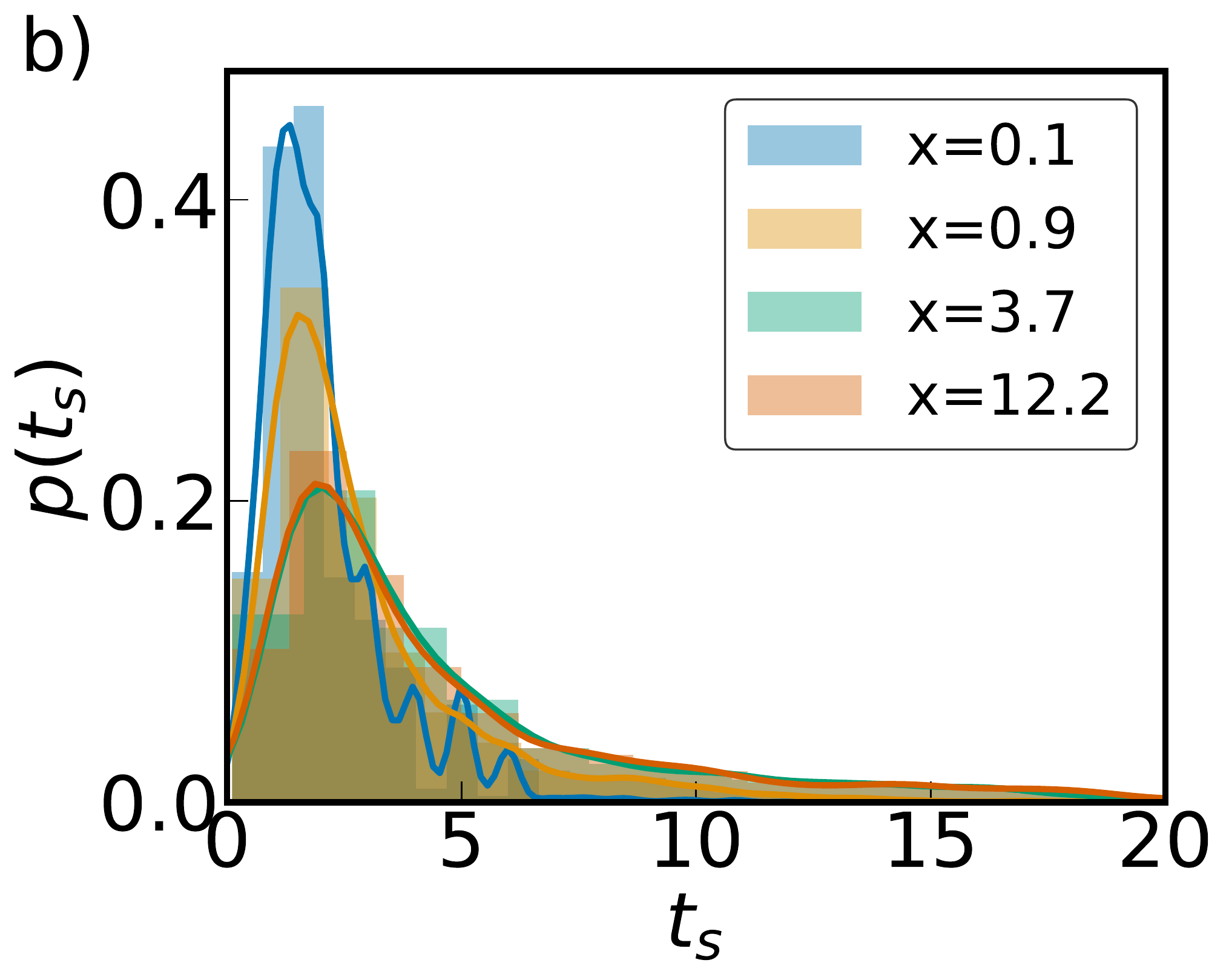}
    \includegraphics[width=0.3\columnwidth]{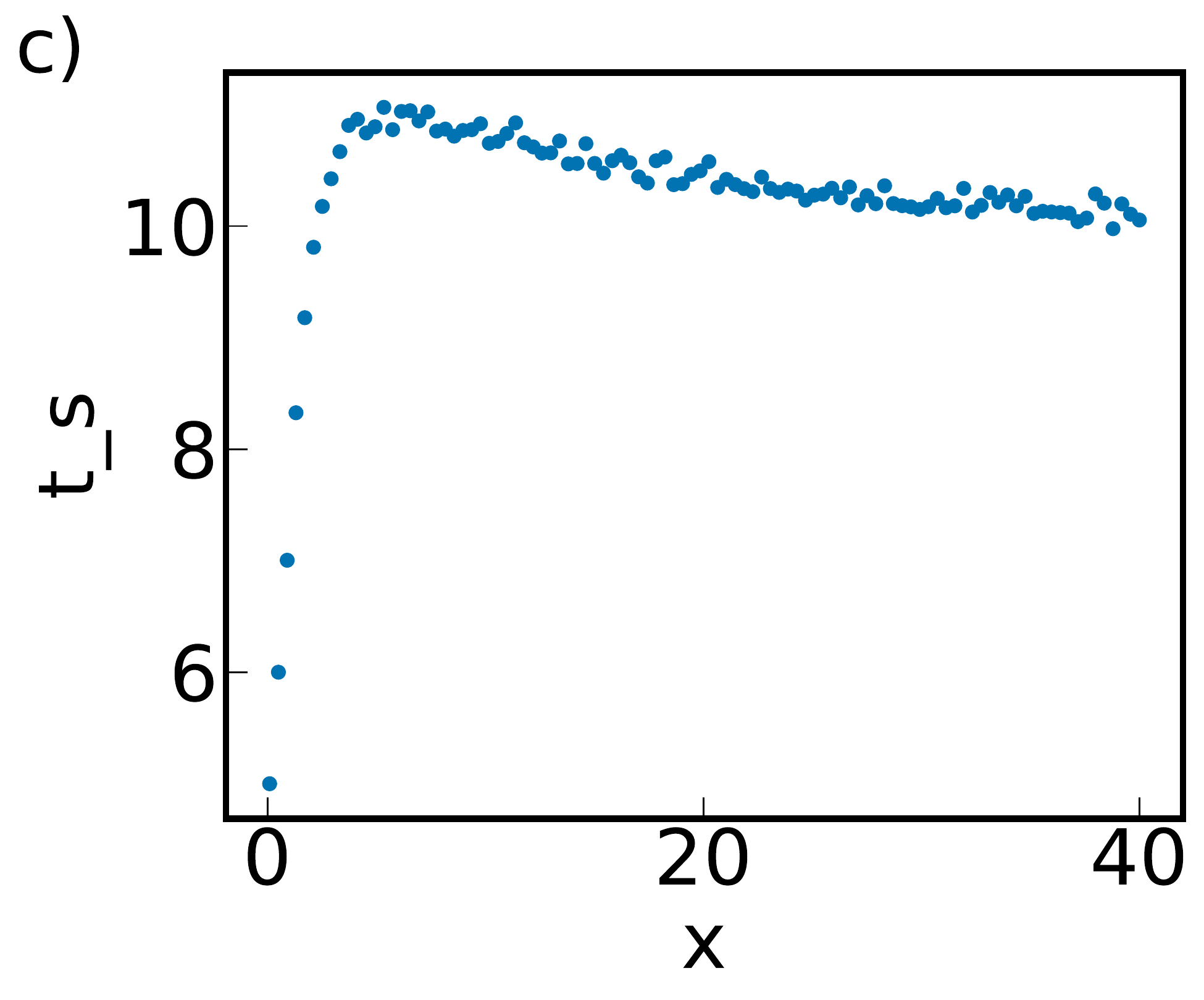}
    \caption{Properties of the route reach an equilibrium state. a) Time series of stop-list-length $n$ for four values of $x$, in a $10$ node ring. Higher request rate $x$ leads to longer stop lists. 
    b) Histogram of (relative) service time $t_s$ for four values of $x$. A higher request rate $x$ leads to a wider distribution of relative service times. c) The mean relative service time grows with the request rate but saturates at the network size $N$, as for large $x$ the bus-route covers the entire network.}
    \label{fig:diff-x-ring}
\end{figure}

Starting from an empty vehicle in a random position, we subsequently generate random requests with pick-up and drop-off nodes chosen uniformly randomly from the nodes of the graph, that are then included in the route by the dispatcher algorithm and served at constant velocity. This is repeated until the steady state is established and performed on a range of different network topologies (star, ring, grid, city layouts) and request rates ($0<x<40$). Results of the simulations on a ring with 10 nodes are shown in Fig.~\ref{fig:diff-x-ring}. The complete simulation code is available in \cite{selfsoftware}.

\section{Analytics}
We analytically derive approximations for the stop list length $n$ in the steady state, by solving the evolution equation for the length of the planned route after $r$ insertions $L_r$. 
\begin{equation}
    L_{r+1}=L_{r}+l^+(n) - \nu \Delta t,
\end{equation}
Where $l^+(n)$ is the added length per request, $\nu \Delta t$ is the distance driven in between requests and the discrete time parameter $r$ counts the requests.

The planned route length reaches its equilibrium at 
\begin{equation}
    l^+(n) = \nu \Delta t =\frac{2\evdel{l}}{x},
    \label{eq.leq}
\end{equation}
where we used the definition of $x$ from Eq.~\ref{eq.x}.

In case of a taxi system the added length is independent of $n$ at two times the average shortest path length in the network
\begin{equation}
    l^+_{taxi}= 2 \evdel{l},
\end{equation}
as new segments are simply added to the end. If $x>1$, the system no longer has an equilibrium as the route length keeps growing, if on the other hand $x<1$, the taxi has time between subsequent rides, in which it stands still, lowering the average velocity.

In a ride-sharing system with a sensible dispatcher algorithm, however the length of the added segments depends on the planned route. In our model for example, as the current length of the route $n$ grows, the probability that a new request's pick-up and/or drop-off being already included in the route increases, resulting in a smaller $l^+(n)$.

We take the added length to be the average over three possibilities:
\begin{enumerate}
    \item [a] Both, pick-up and drop-off node are already on the route.
    \item [b] The pick-up node is on the route but the drop-off node is not.
    \item [c] The pick-up node is not on the route.
\end{enumerate}
\begin{figure}
    \centering
    \includegraphics[width=\columnwidth]{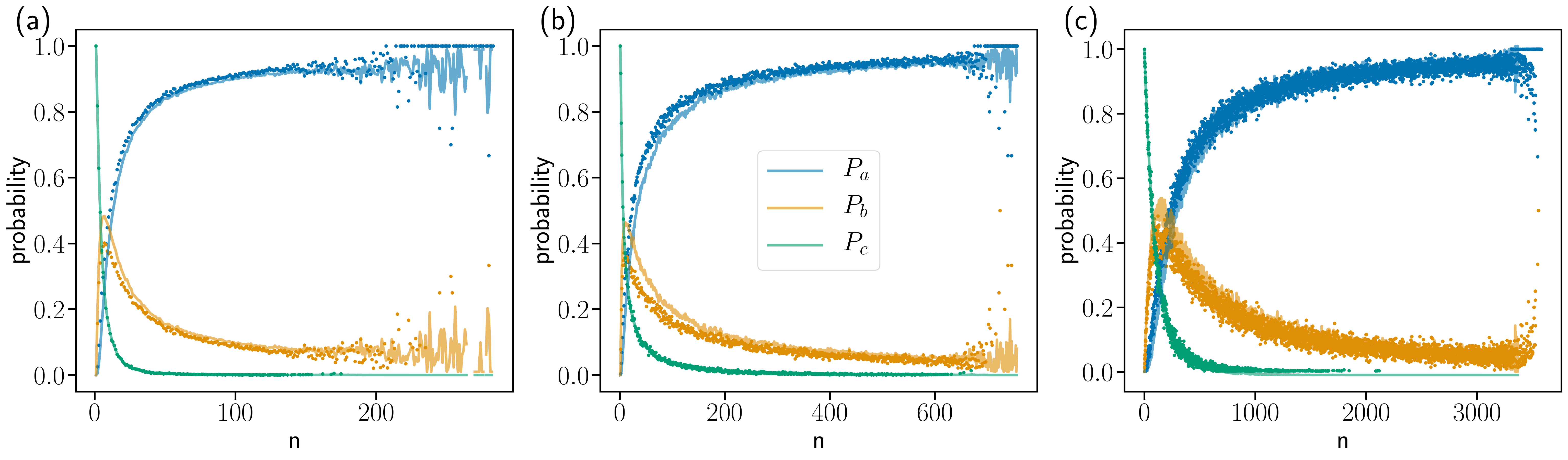}
     \caption{The relative probabilities of insertion types a, b and c (dots) together with the results of Eq.~\ref{eq.Pabc} (lines). a) on a ring network with $N=101$, b) on a square $10 \times 10$ grid, c) on a star graph with $N=100$.}
    \label{fig:pabc}
\end{figure}
In case a) no length is added to the route, in case b) the average added length is $\evdel{l}$ and in case c) the route gets longer by an average of $2\evdel{l}$. This means that
\begin{equation}
    l^+(n) \approx P_b \evdel{l}+2 P_c \evdel{l},
    \label{eq.l+}
\end{equation}
where $P_b$ and $P_c$ are the probabilities of $b$ and $c$ respectively.

\begin{figure}
    \centering
    \includegraphics[height=4.2cm]{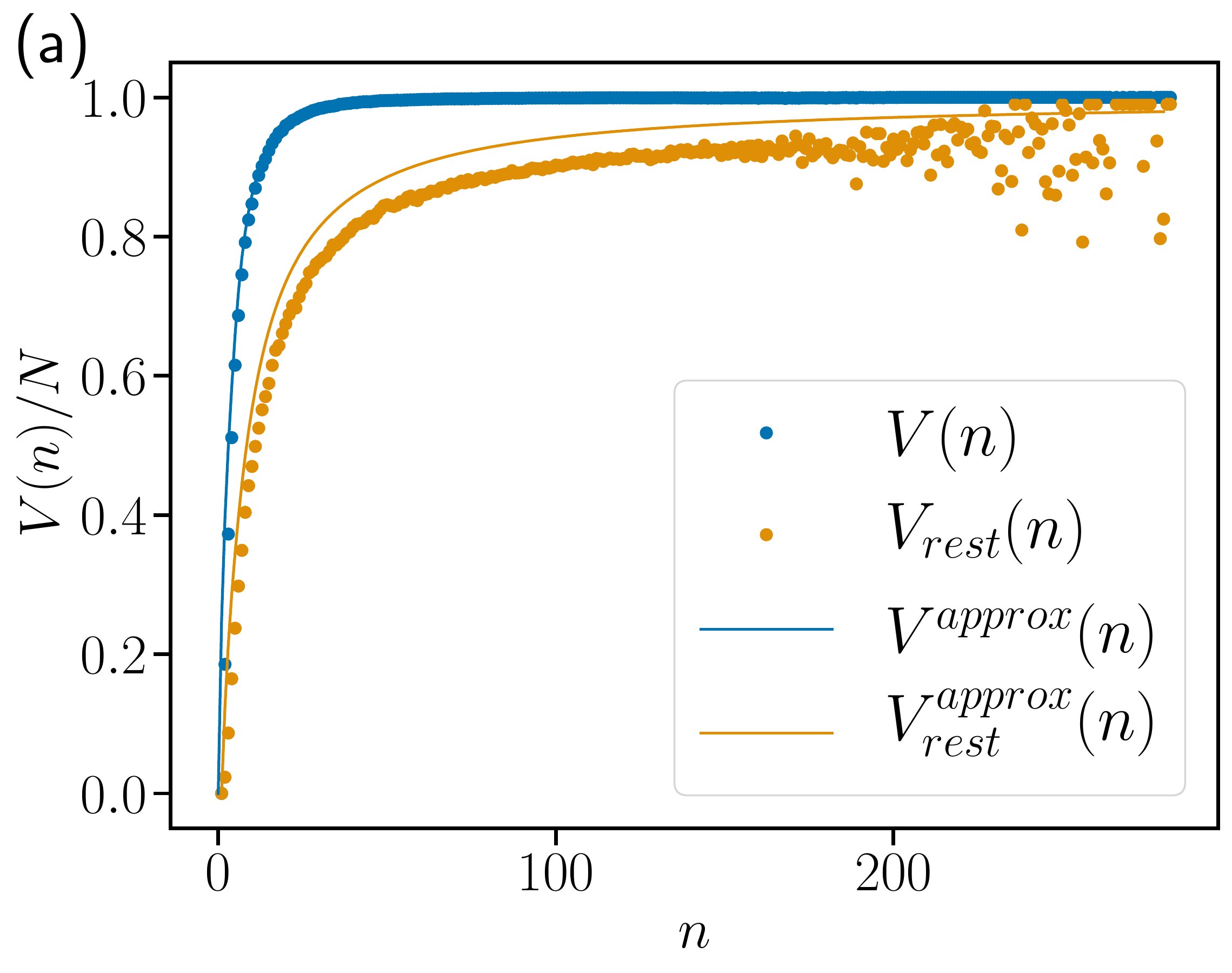}
    \includegraphics[height=4.2cm]{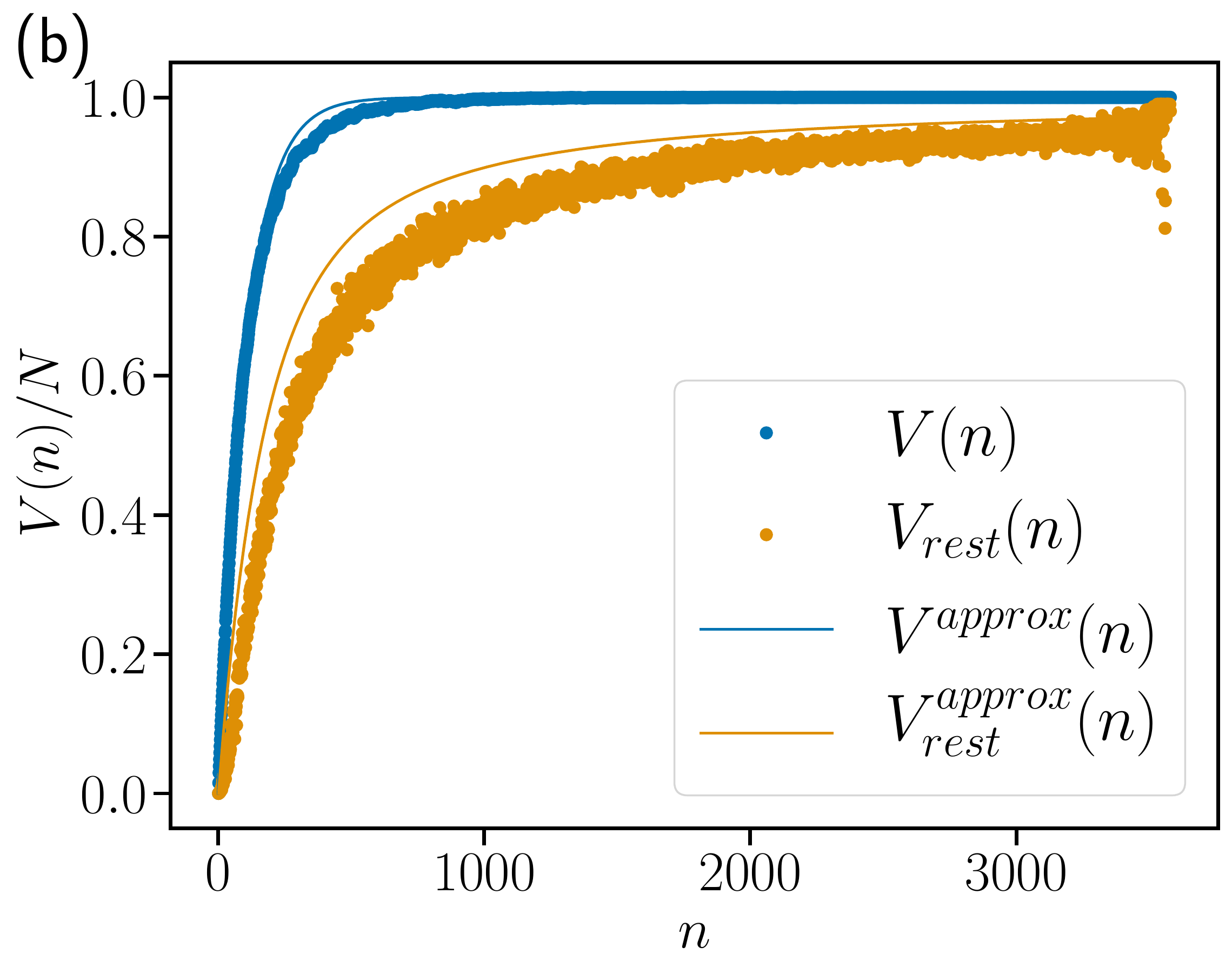}
    \includegraphics[height=4.2cm]{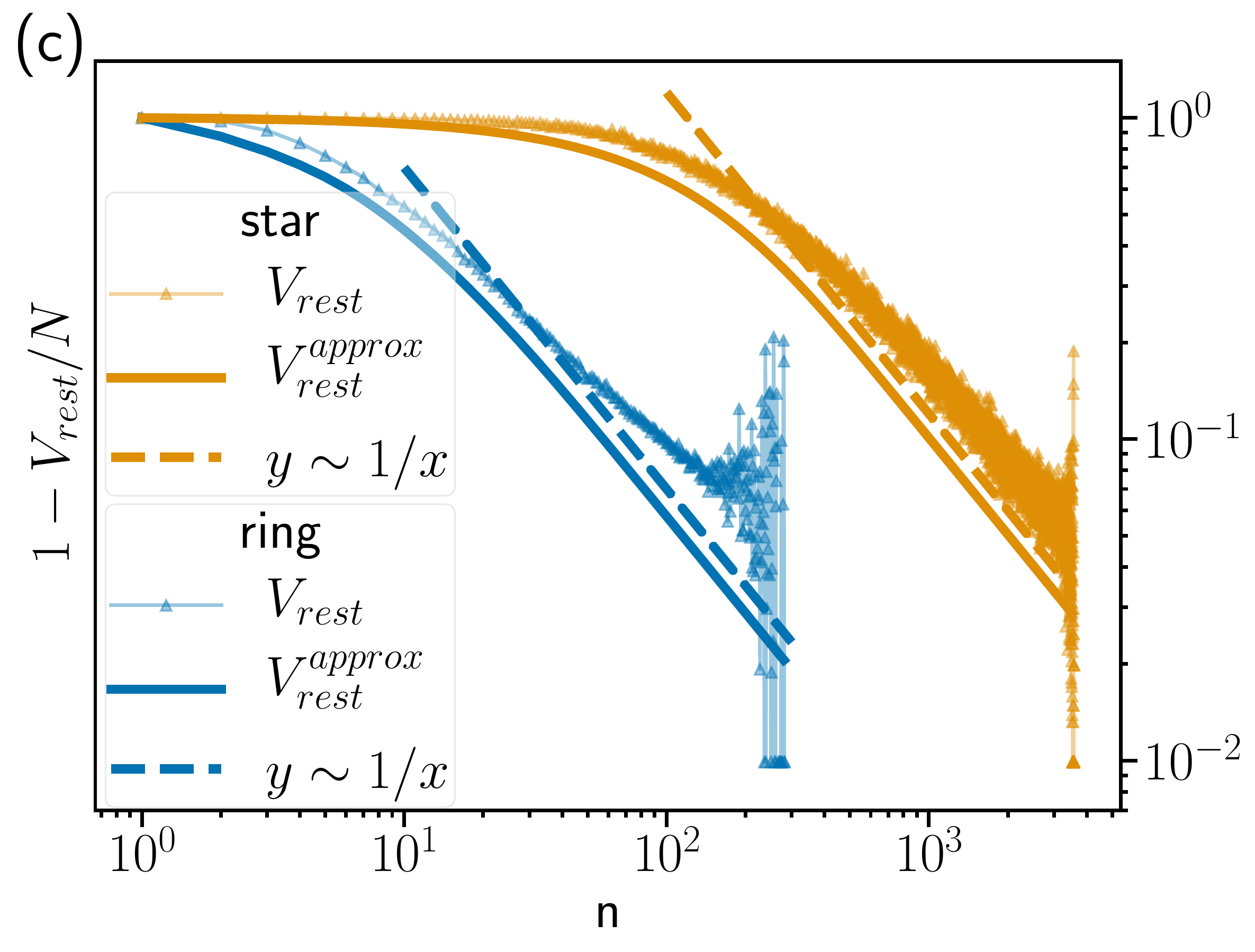}
     \caption{Volume with stop-list length. a) ring graph together with its approximation from Eq.~\ref{eq:dV_main} and b) star graph together with its approximation from Eq.~\ref{eq.vstar}. The analytical approximation for $V_{rest}$ from Eq.~\ref{eq.v2} slightly overestimates the drop-off volume, indicating that the insertions of the pick-ups are not uniformly random across the whole length of the route. c) $V_{rest}$ approaches $N$ indeed as $1/N$ in the asymptotic limit, as predicted in Eq. \eqref{eq.v2_alpha}.}
    \label{fig:V}
\end{figure}

To evaluate the probabilities, we introduce \emph{route volume} $V$ as the number of nodes that can be reached within the stop list without a detour. 
The probability for the requested pick-up node to be on the route depends on the volume $V(n)$ of the route. The volume depends on the length of the stop-list as well as the topology of the underlying network. As the requested drop-off point has to be on the route after the pick-up, there is a second relevant route volume, namely $V_{rest}(n)$, the average volume of the route after the pick-up. Assuming the position of the pick-up is uniformly randomly located somewhere along the route, the fact that the insertion of the drop-off is always after the pick-up leads to (we employ a simplifying assumption that the pick-up is equally likely to be anywhere on the stop-list)
\begin{equation}
    V_{rest}(n) \approx \sum_{k=1}^n V(n-k)/n=\sum_{k=1}^n V(k)/n.
    \label{eq.v2}
\end{equation}
Furthermore, we note that the function $V(n)$ is always monotonously growing with $n$ and asymptotically approaching $N$. We thus express $V_{rest}(n)$ for large $n$ as
\begin{equation}
    V_{rest}(n) = \frac{1}{n} \sum_{k=1}^n \left[N-(N-V(k))\right] = N - \frac{1}{n}\sum_{k=1}^n \left[N-V(k)\right] \approx N\left(1-\frac{\alpha}{n}\right),
    \label{eq.v2_alpha}
\end{equation}
where $\alpha=\sum_{k=1}^n \left[1-V(k)/N\right]$, if this limit exists. Note that $\lim_{n\to\infty}1-V(n)/N=0$, and if in addition we know that $1-V(n)/N$ goes to 0 faster than $1/n$, then $\alpha$ is guaranteed to exist. In this case, $\alpha$ is a constant, depending only on the volume growth in a particular network, so $V_{rest}$ approaches $N$ with $n^{-1}$. We demonstrate in Fig \ref{fig:V} that at least for rings and stars, this assumption holds.

Using this, we can express the probabilities for the three insertion types:
\begin{align}
    P_a &= \frac{V(n)}{N}\frac{V_{rest}(n)}{N} \nonumber \\
    P_b &= \frac{V(n)}{N} (1- \frac{V_{rest}(n)}{N})\nonumber \\
    P_c &= 1- \frac{V(n)}{N}.
    \label{eq.Pabc}
\end{align}
This is shown for a number of different networks in Fig.\,\ref{fig:pabc}.

In a ring of length $N$ the expected route volume $V_{ring}(n)$ for a stop list of length $n$ is given by the recursive relation

\begin{align}
    \label{eq:dV_main}
    V(n+1)  = 
    \begin{cases}
        \frac{N}{4} + \frac{V(n)}{2} + \frac{V(n)^2}{3N}  & \text{ if } V(n) \leq \frac{N}{2} \\
        -\frac{N^2}{4V(n)} + \frac{5N}{4} - \frac{2V(n)}{3} + \frac{2V(n)^2}{3N}& \text{ if } V(n) > \frac{N}{2}.
    \end{cases}
\end{align}

A detailed derivation \eqref{eq:dV} is given in the appendix/supplemental material. This approximation holds very well as shown in Fig.~\ref{fig:V}a.

In a star with $N$ nodes, the number of nodes on the route is approximately equal to the number of unique random draws. This is given by
\begin{equation}
    V_{star}(n)\approx N \left(1-\left(\frac{N-1}{N}\right)^n\right).
    \label{eq.vstar}
\end{equation}
This approximation does not account for the special role of the center point of the star, which is always on the route, as soon as the stop-list contains two or more nodes. Nonetheless it holds reasonably well, as shown in Fig.~\ref{fig:V}b.

\section{Results}
\begin{figure}
    \centering
    \includegraphics[width=0.9\columnwidth]{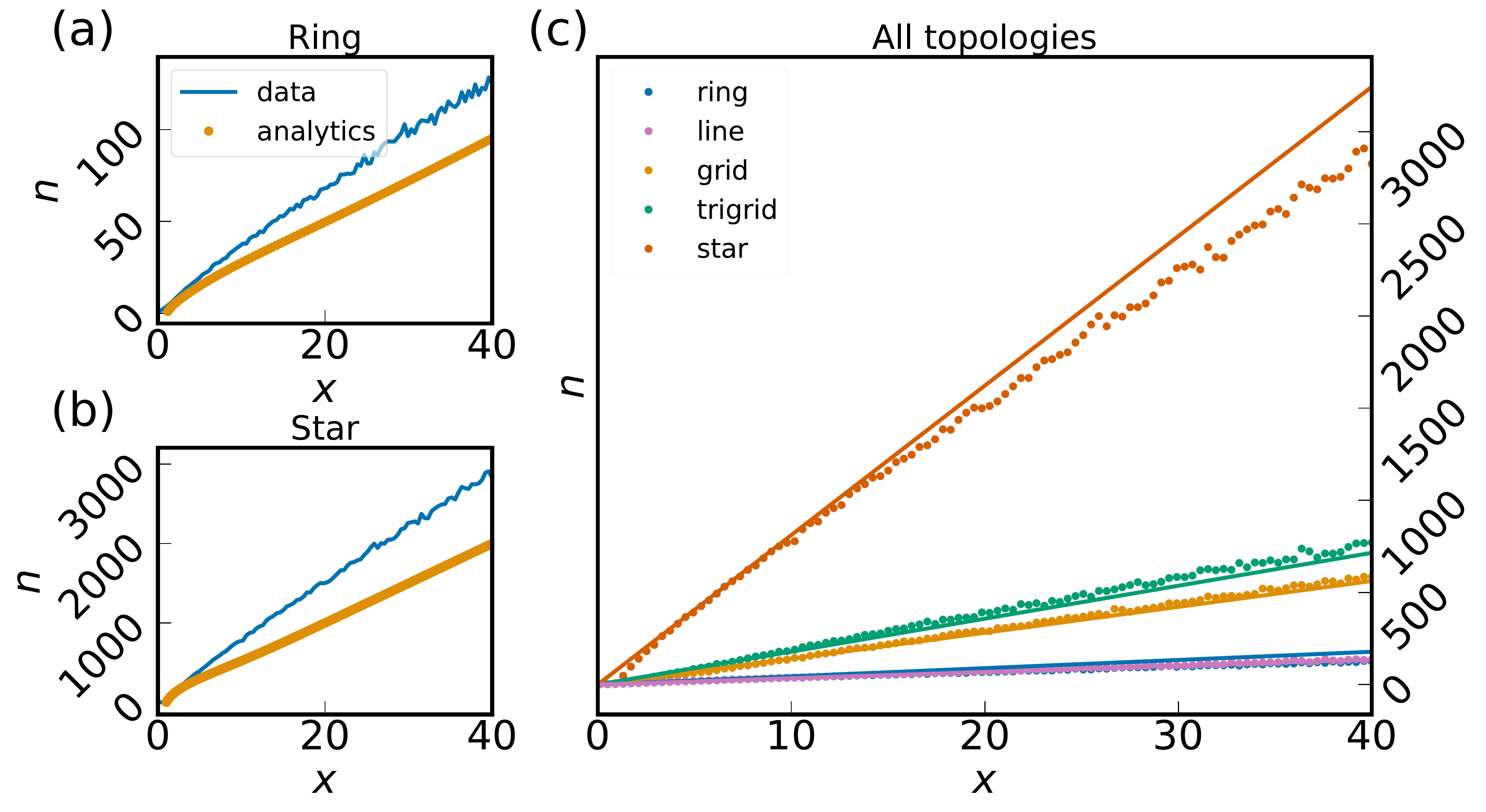}
    \caption{\textbf{Scaling of stop-list length $n$ vs. (dimensionless) request rate $x$.} In the limit for large $x$, $n$ scales approximately linearly with $x$ for all networks. The slope grows with the dimensionality of the street network. (a) 100 node ring, $n$ measured directly (blue line) qualitatively agrees with $n$ reconstructed from volumes using Eq.~\ref{eq:dV_main}(b) 100 node star, $n$ measured directly (blue line) qualitatively agrees with $n$ reconstructed from volumes using Eq.~\ref{eq.vstar} (c) simulated $n$ vs. $x$ for ring, line, grid, triangular grid and star, measured directly (dots) agrees with $n$ reconstructed from measured $V_{rest}$ (line) according to Eq.~\ref{eq.x(n)}.
    }
    \label{fig:n-vs-x}
\end{figure}

We have now gathered all necessary input for computing the expected steady-state stop-list length.
Inserting the approximated volumes from Eq.~\ref{eq:dV_main} and Eq.~\ref{eq.vstar} (or using volumes extracted from the simulation if no such approximation is available) into the probability functions from Eq.~\ref{eq.Pabc} and the approximation of the second volume from Eq.~\ref{eq.v2} to then insert into the steady state added length from Eq.~\ref{eq.l+} and solving for $n$, we find an approximately linear rise of the stop list length with the dimensionless request rate $x$ (see Fig.~\ref{fig:n-vs-x} a) and b)).

Independently of the exact form of $V(n)$ we can exploit the asymptotic behaviour of $V(n)$ and set
$\frac{V}{N}\rightarrow 1$, leading to (using \eqref{eq.leq} and \eqref {eq.l+})

\begin{align}
\frac{2\evdel{l}}{x} = l^+(n) &\approx P_b \evdel{l}+2 P_c \evdel{l} \nonumber \\
&=\evdel{l} \left[\frac{V}{N} \left(1- \frac{V_{rest}}{N}\right) +2 \left(1- \frac{V}{N}\right)\right]\nonumber \\
&=\evdel{l}\left(1-\frac{V_{rest}}{N}\right),
    \label{eq.large_n_limit}
\end{align}

where we inserted the probabilities from Eq.~\ref{eq.Pabc} into Eq.~\ref{eq.l+} and set $\frac{V}{N}\rightarrow 1$.

In this we use the expression for $\alpha$ from Eq.\ref{eq.v2_alpha} and solve for $x$, giving
\begin{equation}
    x\approx \frac{2 n}{ \alpha},
    \label{eq.x(n)}
\end{equation}
where $\alpha=\lim_{n \rightarrow \infty }\sum_{k=1}^n \left[1 - V(k)/N\right]$ is computed from the analytical expressions of $V(k)$ or directly from simulated $V$ and $V_{rest}$. In Fig.~\ref{fig:n-vs-x} the estimated results for $\alpha$ are inserted in Eq.~\ref{eq.x(n)} and plotted with the directly simulated $n(x)$. For each topology and request rate, 10000 requests were simulated, with the origin and destination of each request drawn uniformly randomly from the nodes \cite{selfsoftware}.
% 'ring': 11.386028572570162,
%'line': 6.41897058532523,
% 'star': 186.0432046196512,
% 'grid': 28.24196951524343,
% 'trigrid': 35.45718568776524}

We find values of $\alpha \approx 11.4$ for the ring, $\alpha \approx 6.4$ for the line, $\alpha \approx 28.2$ for the grid, $\alpha \approx 35.4$ for the triangular grid and $\approx 186.0$ for the star.
The resulting lines capture the behaviour of the curves reasonably well.

The star graph has by far the steepest curve, indicating the worst layout for ride-sharing. This was expected as there is only one point that is on the way while all other nodes are detours. The grids perform slightly better, as there are multiple routes between any two points. The ring and line essentially represent the ride-sharing in an elevator, which works without much route adjustment by simply going up and down and collecting whomever is going in the current direction of the elevator.

This shows that ride sharing on a ring or line is very natural and typically possible, while almost no two distinct rides can be bundled on a graph with star topology (see Fig.~\ref{fig:n-vs-x}).

\begin{figure}[!htp]
    \centering
   \includegraphics[width=0.9\columnwidth]{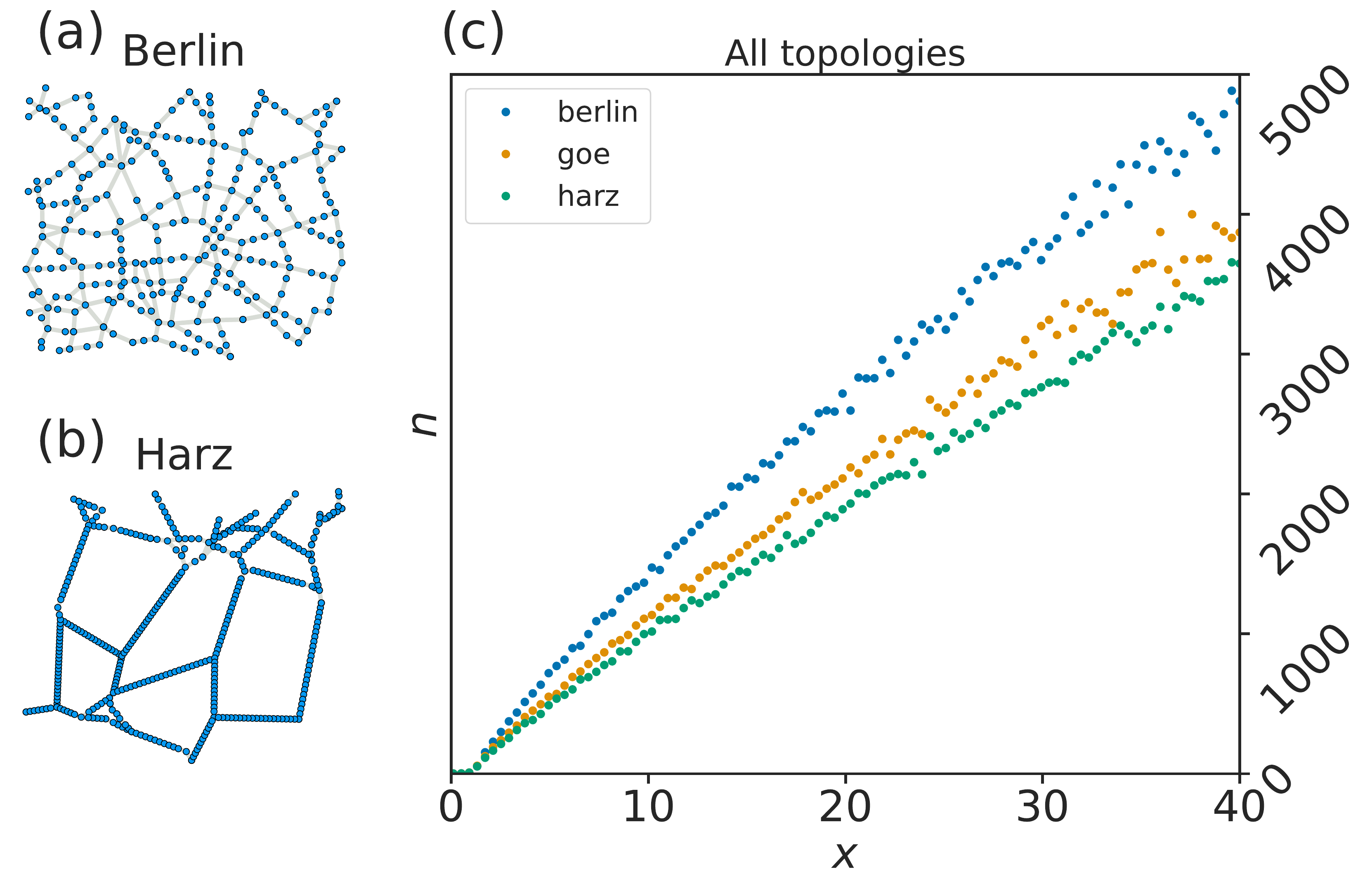}
    \caption{\textbf{Ride-sharing on a variety of street networks.} The length of the planned route $n$ grows linearly with the normalized request rate $x$, with a slope dependant on the topology of the network.
    a) Example street network extracted from Berlin, b) Example street network extracted from Harz, c) $n$ vs. $x$ for networks from Berlin, Göttingen and Harz.}
    \label{fig:data}
\end{figure}
%\subsection{Street networks}

We have applied the methods to real street networks to assess their respective ride-sharing feasibility and compare rural, urban and suburban areas. To this end we have extracted street networks from \texttt{OSMnx} \cite{boeing2017osmnx} and translated the weights into a corresponding number of equally long links, since our method is meant for unweighted graphs. We expect this procedure to have a limited effect on the results as it approximately preserves the lengths of distances. In urban networks, street lengths are largely homogeneous, leading to few added intermediate nodes. In rural areas, on the other hand, street lengths in between settlements are far larger than those within villages, leading to large numbers of added nodes. As a result, street networks in the city, as shown in Fig.~\ref{fig:data} a) resemble a grid, whereas those in the countryside resemble a loose mesh.

We generate networks with $\approx 500$ nodes and simulate 10000 requests in each case. The resulting stop-list length is plotted  in Fig.~\ref{fig:data} over the dimensionless request rate, assuming a uniformly random distribution of requests, as in the case of the artificial networks. We observe linear behaviour with the slope depending on the underlying network, just like we did for synthetic networks in Fig \ref{fig:n-vs-x}.

In particular we find that the grid-like structure of the large city (Berlin) leads to a far steeper slope than the loose mesh of the rural area (Harz), or a single town surrounded by smaller villages (Göttingen), indicating that the rural topology may be more suitable for ride-sharing, when the request pattern is uniformly randomly distributed.

\begin{figure}[!htp]
   \centering
   \includegraphics[width=0.9\columnwidth]{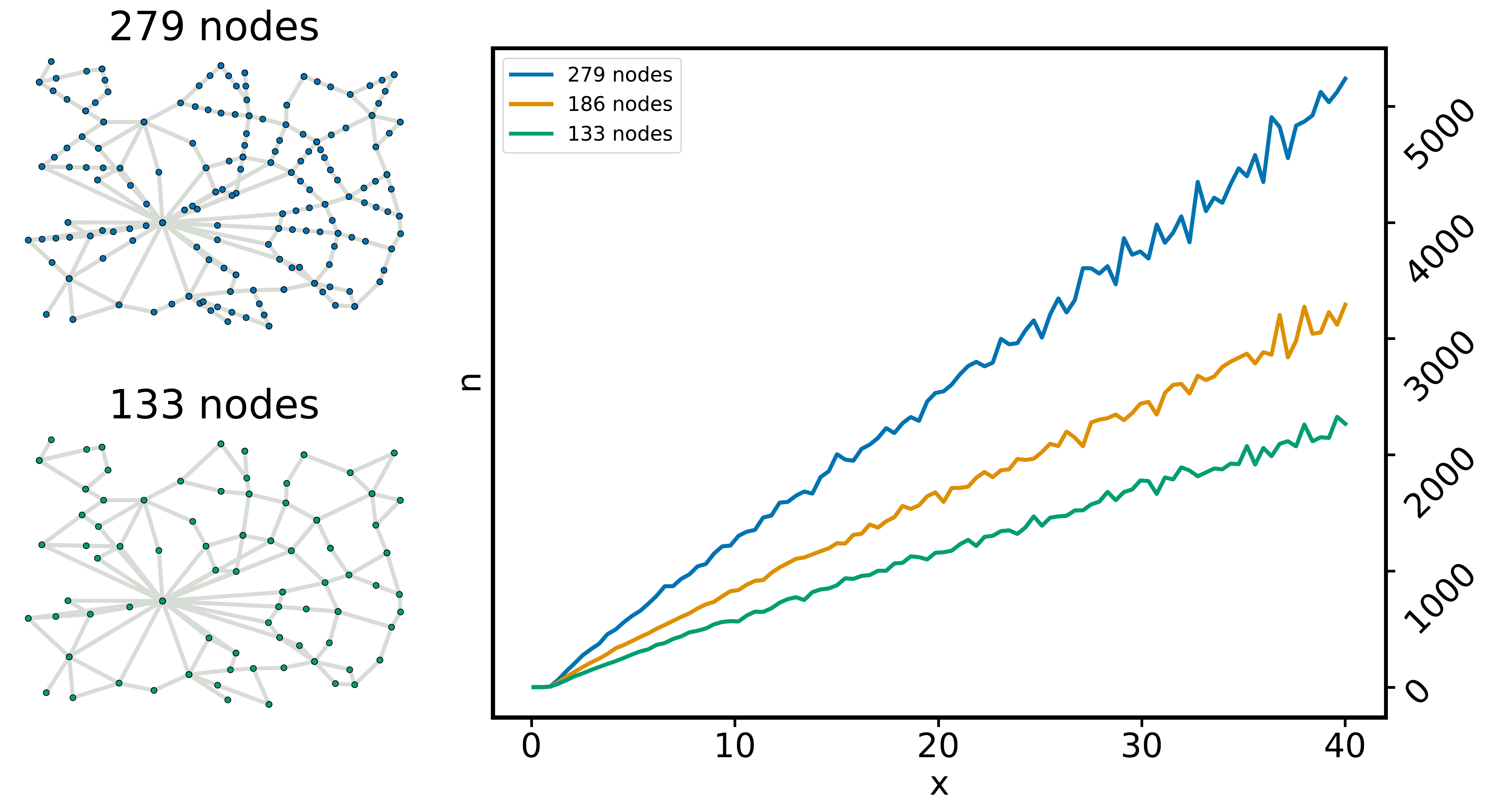}
    \caption{\textbf{Ride-sharing efficiency depends also on network size, in addition to topology.} The slope of $n$ vs. $x$ differs between a few street networks created by differently coarse-graining the street network of Berlin. 
    a) and b) shows two differently coarse-grained and hence, differently-sized networks. c) $n$ vs. $x$ for these two networks. }
    \label{fig:many_berlins}
\end{figure}

Despite the normalization of the request rate with the average shortest path length, we find the slope to depend on the network size, as well as structure, as shown in Fig \ref{fig:many_berlins}.

\section{Conclusion and Discussion}
Here we have tackled the question how network topology affects the feasibility of ride sharing. 
For this, we have studied the steady states of a one-vehicle ride-sharing system in a range of simple homogeneous networks as well as real regional street network topologies using both analytical and numerical methods and analyzed the backlog (i.e. the stop-list) in the system depending on its load.

We find that, while similar networks also result in similar scaling behaviour of the stop-list length (i.e. ride-sharing is almost indistinguishable between the ring and the line or different grids of the same size), there are large differences between networks with different dimensions (i.e. ring (1D), grid (2D) and star ($\infty$D)). Furthermore, there is a substantial impact of the network size $N$ on the ride-sharing predisposition.

We find these differences to be summarized by the cumulative route volume parameter $\alpha$.

In the light of these findings we have compared a range of very different regional street networks ranging from urban centers to smaller towns and rural areas. Assigning them the same request density and selecting bounding boxes, such that the networks have comparable distances and node numbers $N$, we find, 
that ride-sharing is topologically harder in urban areas, as their grid-like structure causes routes to be more likely to be distinct while the loose mesh, characteristic of rural areas topologically forces ride-sharing on long stretches of the connecting streets. We expect this effect to be even stronger with more realistic request patterns, which would remain largely homogeneous in urban centers, but be centered around settlements in rural areas, effectively lowering the number of active nodes.

This effect is, however, counteracted by the typically inconveniently low request rates for public service options in rural areas. To make use of the beneficial network structure, it would therefore be necessary to convince more customers of participating in shared flexible transport options.

The combination of nearby nodes (coarse graining) may improve ride-sharing and thus deliver convenient, efficient public transport. This may be realized in practice by offering the customers a choice of \emph{virtual bus-stops}. 
Cities on the other hand already have an inexpensive and efficient public transport in line-services. Further research is needed to determine how introducing line-services on the most frequented routes would affect or be combined with on-demand ride-sharing.

\section*{Declarations}
\subsection{Availability of data and materials}
All data generated or analysed during this study are included in this published article [and its supplementary information files].
\subsection{Competing interests}
The authors declare no competing interests.
\subsection{Funding}
This research was supported by the European Fond for Regional Development (EFRE) through the state of Lower Saxony, and the Max Planck Society.
\subsection{Authors' contributions}
DM and NM conceived, designed and carried out the research, DM carried out the simulations, DM and NM wrote the manuscript.
\subsection{Acknowledgements}
We thank Stephan Herminghaus, Marc Timme, Malte Schröder, Phillip Marszal, Nils Bayer and Felix Jung for fruitful discussions.

\bibliographystyle{apsrev}
\bibliography{ecobuslit}

\appendix
\section{Stoplist volume calculation}
We want to know what is the expected (i.e. ensemble average) \emph{route 
volume} $V(n)$, when $n$ stops are added to an initially empty stoplist 
according to the procedure described in Section \ref{sec:model}. 
\subsection{Ring}
Consider an $N$ node ring with nodes labelled $1,2,\cdots,N$.
Let a stoplist of length $n$ be constructed as per Section \ref{sec:model} and $\upsilon_n$ be its 
volume. Our goal is to find out the expectation value $E[\upsilon_n]:=V(n)$.

We go to the continuum limit here for ease of analytical computations (i.e. we 
approximate the $N$ node ring with a continuous ring of length $N$). Further, 
we do our calculations in phase variables, by defining
\begin{align}
    \label{}
    \upsilon_n &:= \frac{w_n}{2\pi} N.
\end{align}
We further assume without loss of generality that the stoplist has its 
extremities at $0$ and $\upsilon_n$, effectively going to a rotated coordinate 
system. 

We approach the problem by looking at the expectation value of the 
\emph{increment} in volumes when a node is added to an $n$-length stoplist:
\begin{align}
    w_{n+1} &=w_n + \Delta_n(w_n, \zeta_n, \zeta_{n+1}) \\
    E[w_{n+1}|w_n] &= w_n + E[\Delta_n|w_n].
    \label{eq:def-Delta}
\end{align}

If we know $E[\Delta_n|w_n]$, a recursive formula can be derived for $E[w_n]$.   

\begin{figure}[!htp]
\begin{center}
\includegraphics[width=\columnwidth]{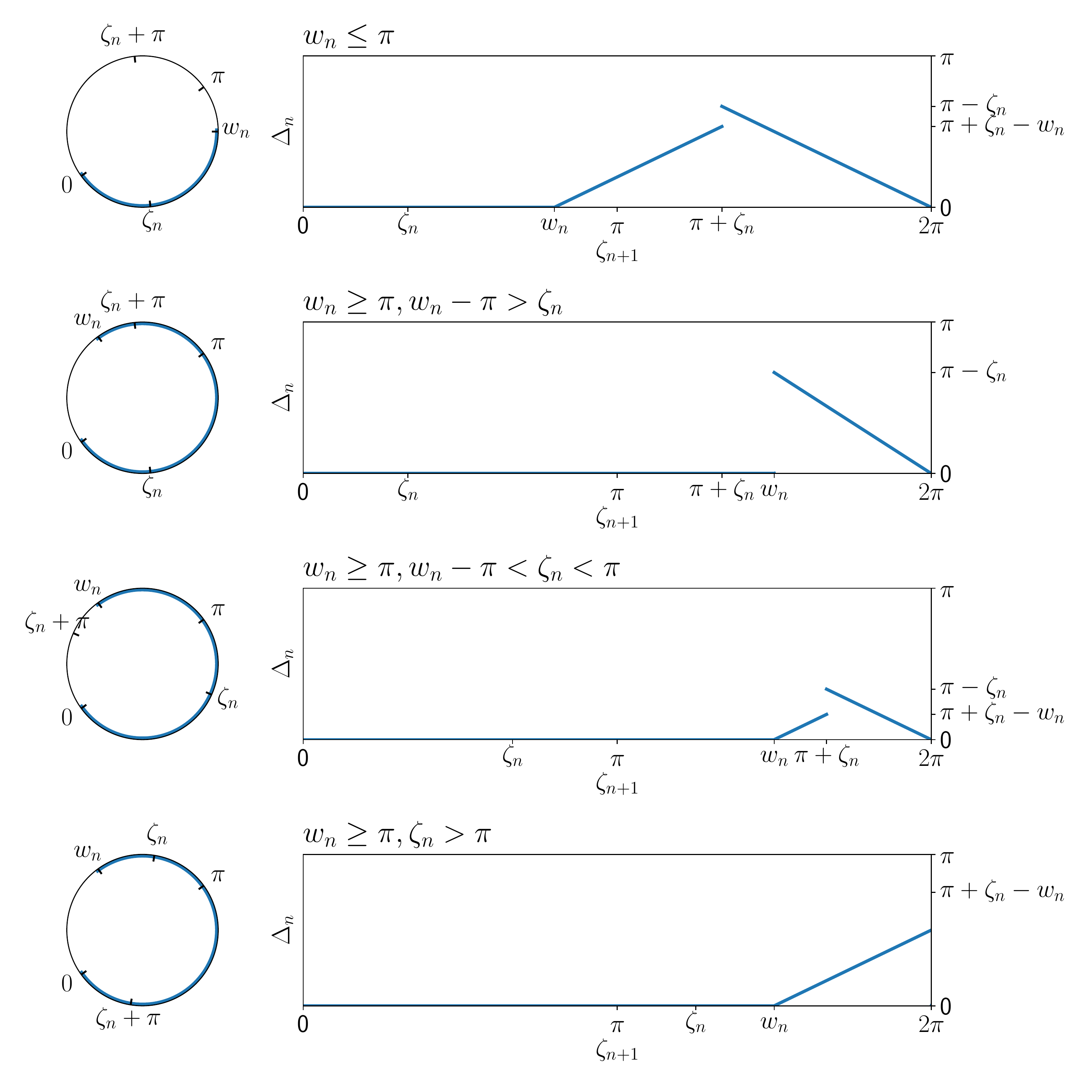}
    \caption{The function $\Delta_n$ as defined in \eqref{eq:def-Delta}}
\label{fig:scaling-nf}
\end{center}
\end{figure}

The stoplist-generation process described in Section \ref{sec:model} necessarily means the 
function $\Delta_n(w_n, \zeta_n, \zeta_{n+1})$ will have the form described in 
Figure \ref{fig:scaling-nf}.

The main difficulty in computing $E[\Delta_n|w_n]$ lies in the fact that although $\zeta_{n+1}$ and $w_n$ are independent, 
$\zeta_n$ and $w_n$ are not independent (because $\zeta_n$, the last point on the stoplist, influences the value of $w_n$). 
We therefore make out first simplifying assumption: Ignore this dependence and assume that $\zeta_n$ is uniformly 
randomly distributed in the interval  
$(0, w_n)$, i.e. the last point in the stoplist is equally likely to be 
found anywhere within the volume. Then we have
\begin{align}
    \zeta_n & \sim U(0, w_n) \\ 
    \zeta_{n+1} &\sim U(0,2\pi),
\end{align}
leading to
\begin{align}
    E[\Delta_n | w_n] &= \int_0^{w_n} d\zeta_n \frac{1}{w_n} 
    \int_0^{2\pi} d\zeta_{n+1} \frac{1}{2\pi} \Delta(w_n, \zeta_n, \zeta_{n+1}).
    \label{eq:baseform}
\end{align}

Plugging in the piecewise-linear functional form of $\Delta_n$ as described in 
Figure \ref{fig:scaling-nf} into Eq. \eqref{eq:baseform} yields
\begin{align}
    \label{eq:delta_w}
    E[\Delta_n|w_n] &= 
    \begin{cases}
        \frac{1}{2\pi} \left(\frac{w_n^2}{3}-w_n\pi+\pi^2\right) & \text{ if } w_n \leq \pi \\
        \frac{2\pi-w_n}{2\pi w_n} \left(-\frac{2w_n^2}{3}+2\pi w_n-\pi^2\right) & \text{ if } w_n > \pi,
    \end{cases}
\end{align}

and using \eqref{eq:def-Delta}, we get
\begin{align}
    \label{eq:w}
    E[w_{n+1}|w_n] &= 
    \begin{cases}
        \frac{\pi}{2} + \frac{w_n}{2} +\frac{w_n^2}{6\pi} & \text{ if } w_n \leq \pi \\
        -\frac{\pi^2}{w_n} + \frac{5\pi}{2} - \frac{2w_n}{3} + \frac{w_n^2}{3\pi}& \text{ if } w_n > \pi.  
    \end{cases}
\end{align}

Now we can go back to the discrete $\upsilon_n$ from the continuous $w_n$ by using Eq 
\eqref{eq:def-Delta}. We also adopt the shorthand
\begin{align}
    V(n) &:= E[\upsilon(n)],
\end{align}
resulting in 
\begin{align}
    \label{eq:dV}
    V(n+1)  = 
    \begin{cases}
        \frac{N}{4} + \frac{V(n)}{2} + \frac{V(n)^2}{3N}  & \text{ if } V(n) \leq \frac{N}{2} \\
        -\frac{N^2}{4V(n)} + \frac{5N}{4} - \frac{2V(n)}{3} + \frac{2V(n)^2}{3N}& \text{ if } V(n) > \frac{N}{2}.
    \end{cases}
\end{align}

\end{document}